\documentclass[twocolumn,nofootinbib,showpacs,preprintnumbers,amsmath,amssymb,prd,floatfix]{revtex4}
\usepackage{epsfig}
\usepackage{dcolumn}% Align table columns on decimal point
\usepackage{bm}% bold math
\usepackage{natbib}
\usepackage{verbatim}
\usepackage{latexsym}
\usepackage{texdraw}
\usepackage{color}
\usepackage{epstopdf}
\usepackage{hyperref}

\newcommand{\update}[1]{{\color{black}{#1}}}

\begin{document}
\newcommand{\qm}{\frac{q}{m}}
\title{SNELL'S LAW AND REFRACTION OF ELECTRON WORLD LINES BY INTENSE LASER FIELDS}
%\title{REFRACTION AND  SNELL'S LAW FOR ELECTRON DYNAMICS IN INTENSE FEW-CYCLE LASER FIELDS}
\author{Ulrich H. Gerlach}\email{gerlach@math.ohio-state.edu}
\affiliation{%
Department of Mathematics, Ohio State University, Columbus, OH 43210, USA}
\author{Linn Van Woerkom}\email{lvw@mps.ohio-state.edu}
\affiliation{Department of Physics, Ohio State University, Columbus, OH 43210 USA}
\author{ Kiam H. Kwa}\email{khkwa@math.ohio-state.edu
}
\affiliation{Institute of Mathematical Sciences, Faculty of Science, University of Malaya, 50603 Kuala Lumpur, Malaysia}
\date{\today}% It is always \today, today,
             %  but cdany date may be explicitly specified
\begin{abstract}
The dynamics of an electron driven by arbitrary plane wave laser
radiation is reformulated as a relativistically and mathematically exact
refraction process based on an exact index of refraction.  \update{This
reformulation leads to that index as an indicator of the energy
transfer between the electron and the radiation.} It also leads
to the dynamics of the electron as being governed (i) by the
Lorentzian version of what in Euclidean space is Snell's law, \update{(ii) by an
eikonal equation with the corresponding index of refraction, (iii)
by geodesics on a spacetime manifold with in-general non-zero curvature
(``gemetrization of the laser radiation''),} (iv) by the spacetime version
of Fermat's principle of least time, and (v) by a Lorentzian stability
criterion for the circumstance which in Euclidean space corresponds to
the propagation of rays passing through a periodic wave guide of
lenses which all have the same focal length. That criterion demands
that the laser intensity satisfy $I_{aver}<.56 \cdot
(10,000$\AA$/\lambda)^2 \times10^{18}$watts/cm$^2$.
\end{abstract}
\maketitle
%\section

%\centerline{\bf I. INTRODUCTION:\\ THREE LAWS OF LASER-DRIVEN CHARGE DYNAMICS}

When a particle of charge $q$ and mass $m$ is placed into a laser beam
whose radiation frequency $\omega /2 \pi$ has electric field amplitude
$E_0$, then the particle executes oscillatory motion. The magnitude of
the this effect is expressed by the dimensionless \emph{impulse factor}
\[
\frac{qE_0}{mc\omega}\equiv \eta ~.
\]
In light of (a) the simultaneous presence of an oscillating magnetic
field, and (b) the possibility of the motion being relativistic, it is
not surprising that the resulting complexity in the actual motion of
the particle implies a corresponding complexity of the mathematical
description.

However, it has turned out that, hidden behind this complexity, there
often exists a readily identifiable simplicity, which physicists have
expressed in terms of what is known as the ``ponderomotive potential''
and its gradient, the ``ponderomotive force''. These quantities arise
from the fact that quite often the full motion of a particle is characterized by two
time scales.  One characterizes the rapid quivering/oscillatory (fine-grained)
aspect of the full motion. The other characterizes a slower ``guiding
center'' (coarse-grained) motion around which the particle executes
its fast quivering
oscillations\cite{LandauMechanics,Kibble:1966a,Kibble:1966b,Bauer:1995,Startsev:1997}.

The ponderomotive force/potential, a slowly varying function of space
and time, determines the slow motion dynamics of the particle. It is
the result of performing a one-cycle average over the complex motion
of the particle. There are a number of
ways\cite{Kibble:1966a,Kibble:1966b,Bauer:1995,Startsev:1997}
of doing this, but their common drawback is that one not only loses
potentially useful information about the particle's motion but, more
importantly, misrepresents it, when the laser radiation becomes so
intense ($I_{aver}\cdot(\lambda/10,000\textrm{\AA})^2\ge 1.37 \times
10^{18}~\textrm{watts/cm}^2~\Leftrightarrow ~\eta\ge 1$) that the
ponderomotive force varies as rapidly as the one due to the ``rapid''
quiverings/oscillations.  Under such a circumstance, which includes a
charge in a standing plane wave\cite{Bauer:1995,
Kaplan:2005,Pokrovsky:2005}, an assumed decomposition into
oscillatory plus averaged motion along the direction of the laser
beam, does not apply.(See, however, the second footnote below). For
one thing, an \emph{ab initio} averaging hides the possibility of
resonance\cite{ArnoldAvez:1968} where the frequency of the oscillatory
motion is a multiple of that of the averaged motion.

Furthermore, femtosecond pulsed laser radiation makes any averaging
scheme meaningless. There is not enough time to establish an average
which changes slowly.  Neither does averaging apply to charges located
in, and hence scattered by, the transient overlap region of two
counter propagating few cycle pulses.

There is a superior way of understanding the dynamics of an electron
driven by arbitrary plane wave laser radiation. We shall introduce an
\emph{index of refraction} which is \emph{exact} for arbitrarily
relativistic motion and/or arbitrarily short laser pulses and then
reformulate the dynamics as a Lorentzian refraction process, as
compared to its well known Euclidean cousin. Taking a cue from the
principles of geometrical optics, this reformulation leads to (i) the
Lorentzian version of what in Euclidean space is Snell's law, (ii) the
spacetime version of Fermat's principle of least time, (iii) a
Lorentzian stability criterion for the circumstance which in Euclidean
space is the propagation of rays passing through a periodic wave guide
of lenses which all have the same focal length. Although they are
cousins, the Lorentz and the Euclidean refraction processes are
fundamentally distinct. Their respective arenas, Euclidean space and
space-time, entail (as shown below) opposite ray behavior, attraction
for one, repulsion for the other. This difference can be traced to the
difference in sign in the two respective Fermat variational principles
exhibited below.
 
\emph{Dynamics of a Charge in a Generic Plane Wave Field} --
Consider an electron born at a generic location of the
electromagnetic plane wave field of a laser directed along the
$z$-direction. The four components of the vector potential are
\begin{align}
\{ A_0,A_1,A_2,A_3\}=\{0,0,A_x(t,z),A_y(t,z)\}~.
\label{eq:Vetor potential for laser only}
\end{align}
The dynamics of the particle is governed by the four Lorentz (force
law) equations of motion\cite{MTW:3.1,Jackson:1975} $md^2x^\alpha
/d\tau^2=q F^\alpha_{~\beta} dx^\beta /d\tau$ ($\alpha=0,1,2,3$) for
an electron of mass $m$ and charge $q$ in the
e.m. field $F_{\alpha \beta}=\partial_\alpha A_\beta -\partial_\beta
A_\alpha$. For a generic plane wave along the $z$-direction, these
equations imply
\begin{align}
m\frac{d^2t}{d\tau^2}&=+\frac{\partial}{\partial t}\Phi(t,z)
\label{eq:tt-equation}\\
m\frac{d^2z}{d\tau^2}&=-\frac{\partial}{\partial z}\Phi(t,z)
\label{eq:zz-equation}\\
m\frac{dy}{d\tau}&=p_y-q A_y(t,z)\label{eq:yy-equation}\\
m\frac{dx}{d\tau}&=p_x-q A_x(t,z)\label{eq:xx-equation}.
\end{align}
They are invariant under gauge changes having plane wave symmetry.
Here time is measured in light traveling distance, 
$(p_x,p_y)$ is the particle's conserved (``canonical'') transverse momentum,
and
\begin{equation}
\Phi(t,z)=\frac{m}{2}\left\{ \left(\frac{p_x}{m}-\qm A_x(t,z)\right)^2+ 
          \left(\frac{p_y}{m}-\qm A_y(t,z)\right)^2\right\}~,
\label{eq:scalar potential}
\end{equation}
is the scalar potential which characterizes the longitudinal dynamics
in the $(t,z)$-plane. 

The dynamics of the laser-accelerated charge is controlled entirely by
Eqs.(\ref{eq:tt-equation}) and (\ref{eq:zz-equation}).  We therefore
refer to them as the \emph{master system of equations}.  By contrast,
Eqs. (\ref{eq:yy-equation}) and (\ref{eq:xx-equation}) are merely
\emph{slave equations}.  In relation to the longitudinal degrees of
freedom these equations are dynamically passive. They have no effect
on the particle dynamics in the $(t,z)$-plane. Instead, they identify
a physical measurable property, the transverse $x$ and
$y$-velocity components.

The master system of equations implies that 
\begin{equation}
\frac{1}{2}m\left( \frac{dt}{d\tau}\right)^2 -\frac{1}{2}m\left( \frac{dz}{d\tau}\right)^2
-\Phi(t,z)\equiv\mathcal{H}=\frac{m}{2}\label{eq:Hamiltonian}
\end{equation}
is an integral of motion.  The integration constant $\frac{m}{2}$ has
been chosen such that the laboratory clock and a clock co-moving with
the particle remain synchronized whenever the particle is at rest
$\left( \frac{dz}{d\tau}= \frac{dx}{d\tau}=\frac{dy}{d\tau}=0\right)$
in the lab frame.

%%%%%%% added March 14, 2013 %%%%%%%%%%%%

\update{
\emph{Dynamics Mathematized in Terms of Geometry} -- The two master
equations lend themselves to being blended into a new mental unit, a
geodesic on a Lorentzian manifold with a metric whose curvature is a 
geometrization of the electromagnetic field of the laser. This blending 
process is achieved in two steps:
\begin{enumerate}
\item
Note that these master equations,
\begin{align}
m\frac{d^2x^A}{d\tau^2} + \eta^{AB}\frac{\partial\Phi}{\partial x^B}=0;~~\{x^A:x^0=t,x^1=z\}
\label{eq:master equations}
\end{align} 
follow from the standard dynamical variational principle 
\begin{align}
I&=\int\left\{ \frac{m}{2} \eta_{\mu\nu}\frac{dx^\mu}{d\tau}\frac{dx^\nu}{d\tau}
           +qA_\mu\frac{dx^\mu}{d\tau}\right\}d\tau\\
&=\int\left\{ \frac{m}{2} \eta_{AB}\frac{dx^A}{d\tau}\frac{dx^B}{d\tau}+
           \frac{m}{2} \eta_{ab}\frac{dx^a}{d\tau}\frac{dx^b}{d\tau}
           +qA_b\frac{dx^b}{d\tau}\right\}d\tau.
\end{align}
Here upper case and lower case indices refer to the longitudinal
$\{x^A:x^0=t,x^1=z\}$ and transverse coordinates $\{x^a:x^2=x,x^3=y\}$
coordinates respectively.
\item
With the introduction of the transverse components Eq.(\ref{eq:yy-equation})-(\ref{eq:xx-equation}) and the addition of the inconsequential term $-p_a\frac{dx^a}{d\tau}$ the variational principle for the master equations becomes
\begin{align}
I=\int\left\{ \frac{m}{2} \eta_{AB}\frac{dx^A}{d\tau}\frac{dx^B}{d\tau}-
           \Phi\right\}d\tau.
\end{align}
\end{enumerate}
In this variational principle with its two degrees of freedom, the conserved transverse momentum components, $\{p_a:p_x,p_y\}$, are merely parameters, the
transverse momentum components of the particle. The  dynamics of this system is governed by its Hamiltonian-Jacobi (H-J) equation
\begin{align}
\frac{1}{2m}\eta^{AB}\frac{\partial S}{dx^A}\frac{\partial S}{dx^B}+\Phi=-\frac{m}{2}~.
\end{align}
In geometrical optics the equivalent equation 
\begin{align}
\eta^{AB}\frac{\partial S}{dx^A}\frac{\partial S}{dx^B}=-m^2n^2~,
\end{align}
with $n^2$ given by Eq.(\ref{eq:refractive index}) below, is called the \emph{eikonal equation} \cite{Wolf:1980}
having index of refraction $n$. 

The dynamics of our system on the 2-d (flat) Minkowski spacetime is a study
of those particles having rest mass which \emph{varies} with spacetime
location as specified by the e.m. field on the right side of the H-J
equation.  However, such a study is mathematically equivalent
to the one of particles with \emph{fixed} rest mass but on a manifold with a
metric which is location dependent.  This statement is expressed
mathematically by the H-J equation
\begin{align}
g^{AB}\frac{\partial S}{dx^A}\frac{\partial S}{dx^B}+m^2=0
\end{align}
whose inverse metric
\begin{align}
g^{AB}=\frac{\eta^{AB}}{1+\eta^{ab}\left(\frac{p_a}{m}-\frac{q}{m}A_a\right)
                               \left(\frac{p_b}{m}-\frac{q}{m}A_b\right) }
\label{eq:inverse metric}
\end{align}
is location dependent, and whose corresponding metric is 
\cite{Kwa:2009, Kwa:2012}
\begin{align}
g_{AB}&=\eta_{AB}
        \left[1+\eta^{ab}\left(\frac{p_a}{m}-\frac{q}{m}A_a\right)
                         \left(\frac{p_b}{m}-\frac{q}{m}A_b\right) \right]\\
&\equiv \eta_{AB}e^{2\sigma}~.
\end{align}
Its curvature is \cite{Kwa:2009, Kwa:2012}
\begin{align}
R^A_{~~BCD}=\mathcal{R}\left( \delta^A_C \, \eta_{BD}-\delta^A_D\,\eta_{BC}\right)\label{curvature}
\end{align}
This means, among other things, that the curvature vanishes whenever the laser radiation $\{ A_x,A_y\}$ is such as to satisfy 
\[
\mathcal{R}\equiv \frac{\partial^2\sigma}{\partial t^2}- 
                        \frac{\partial^2\sigma}{\partial z^2}=   0~,
\]
which is to say, the radiation propagates into the $+z$ or the $-z$
direction, while any mixture of the two will result in non-zero
curvature.  In the first case the system is integrable; the H-J
equation can be solved by the method of separation of variables. In
the second case the laser-particle system is one which is
non-integrable.

\emph{Laser Driven Dynamics in a Longitudinal Electric Field} A grasp
of laser-induced charge dynamics on a fundamental level requires
taking cognizance of strong Coulomb (``space charge'') fields. Although their forces are
strictly longitudinal, their indirect cause is the strong charge
separation due to the ultra-intense laser interaction with the target
environment \cite{Brunel:1987,Brunel:1988,Geidre:2010,Velcheva:2012}.

With such fields present, instead of 
     Eq.(\ref{eq:Vetor potential for laser only}), the components of the vector potential are
\begin{align}
\{ A_0,A_1,A_2,A_3\}=\{&A_t(t,z),A_z(t,z),A_x(t,z),A_y(t,z)\}~,\\
\intertext{while the master Eqs.(\ref{eq:master equations}) with its additional Coulomb field}\nonumber \\
F_{BC}&=\partial_BA_C-\partial_CA_B\\
\intertext{are}
m\frac{d^2x^A}{d\tau^2} + \eta^{AB}\frac{\partial\Phi}{\partial x^B}&=q\eta^{AB}F_{BC}\frac{dx^C}{d\tau}~.
\end{align}
The reduced variational principle giving rise to them is
\begin{align}
I=\int\left\{ \frac{m}{2} \eta_{AB}\frac{dx^A}{d\tau}\frac{dx^B}{d\tau}+
qA_B\frac{dx^B}{d\tau} -  \Phi\right\}d\tau.
\end{align}
and the corresponding geometrized H-J equation is 
\begin{align}
g^{AB}\left(\frac{\partial S}{dx^A}-qA_A\right) 
                  \left(\frac{\partial S}{dx^B}-qA_B\right)+m^2=0
\end{align}
Its inverse metric is still given by Eq.(\ref{eq:inverse metric}).
 
  }

\emph{Dynamics as Geometrical Optics} -- The integral of motion,
Eq. (\ref{eq:Hamiltonian}), has a property that leads directly to a
geometrical optics formulation of the dynamics of the particle:
Rewrite the integral in the form
\begin{align}
\left( \frac{dt}{d\tau} \right)^2 -\left( \frac{dz}{d\tau} \right)^2 =n^2(t,z)
\label{eq:integral in terms of refractive index}
\end{align}
where
\begin{align}
n(t,z)=\sqrt{1+\left(\frac{p_x}{m}-\qm A_x(t,z)\right)^2+  \left(\frac{p_y}{m}-\qm A_y(t,z)\right)^2}
\label{eq:refractive index}
\end{align}
The integral, Eq.(\ref{eq:integral in terms of refractive index}),
holds along every spacetime trajectory. This suggests that one should
replace the proper time $\tau$ of the charge with its
\emph{longitudinal} proper time\footnote{This time is measured by a
  clock moving along fixed $x$ and $y$ coordinates, but co-moving (and
  accelerating) with the charge strictly along the $z$-direction.}
\begin{align}
\int d\tilde{\tau}=\int n\left( t(\tau),z(\tau)\right)d\tau
\end{align}
as the world line parameter. Such a replacement yields
%\begin{align}
$\frac{d\tilde{\tau}}{d\tau}=n~(>0)$.
%\end{align}
Consequently, the integral of motion becomes
\begin{align}
\left( \frac{dt}{d\tilde\tau} \right)^2 -\left( \frac{dz}{d\tilde\tau}
\right)^2 =1
\label{eq:integral in terms of longitudinal time}
\end{align}
and Eqs. (\ref{eq:tt-equation})-(\ref{eq:zz-equation}) become
the \emph{optical master equations}
\begin{align}
\frac{d}{d\tilde\tau} n \frac{dt}{d\tilde\tau}&=+\frac{\partial
  n}{\partial t}\label{eq:ttt-equation}\\
\frac{d}{d\tilde\tau} n \frac{dz}{d\tilde\tau}&=-\frac{\partial
  n}{\partial z}~,\label{eq:zzz-equation}\\
\intertext{or more succinctly,}
\frac{d}{d\tilde\tau} n \frac{dx^A}{d\tilde\tau}&=\eta^{AB}\frac{\partial
  n}{\partial x^B}~;~~\eta^{AB}=\left[ \begin{array}{cc}
                                          1&0\\
                                          0&-1
                                       \end{array} \right]~;~~
                                       \left. \begin{array}{c}
                                         A\\
                                         B
                                         \end{array}\right\}=0,1~.
\intertext{Compare these equations with those for a light ray through
  a medium with refractive index $n(x^1,x^2,x^3)$ in geometrical
  optics\cite{BornAndWolf:1980},} 
\frac{d}{d s} n
\frac{dx^i}{ds}&=\delta^{ij}\frac{\partial n}{\partial x^j}~;~~ 
                                  \left. \begin{array}{c}
                                    i\\
                                    j
                                    \end{array}\right\}=1,2,3 ~,
\end{align}
where $\delta^{ij}$ is the Kronecker delta, a sum over repeated indices
is understood, and $s$ is the geodesic
length parameter along the ray.  One sees that if $n(x^1,x^2,x^3)$ is
the familiar Euclidean index of refraction for a medium in Euclidean
space, then Eq.(\ref{eq:refractive index}) is the \emph{Lorentzian
  index of refraction} for the e.m.-induced medium in Lorentzian
spacetime. The former determines the ray trajectories $x^i(s)$; the
latter determines the world lines $x^A(\tilde{\tau})$.
%%%%%%%%%%%%%%%%%%%%%%%%%%%%%%%%%%%%%
\begin{figure}[h!]
\centering
\epsfig{file=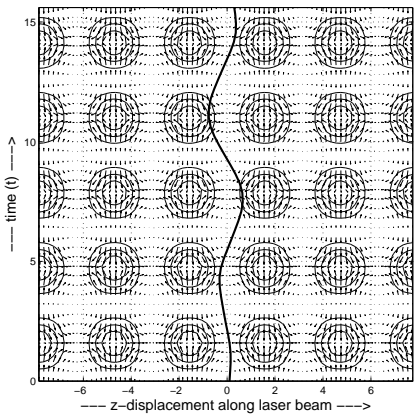,scale=.4}
\caption{World line of a charge through the periodic spacetime lattice
of the electromagnetic standing wave field of a laser. The periodic
index of refraction, Eq.(\ref{eq:squared index for a standing wave}),
has maxima which are surrounded by the oval shaped isograms.  The
gradient vector field is (Lorentz) orthogonal to these
isograms. Furthermore, this vector field pushes the charge and thereby
imparts to it the acceleration given by
Eq.(\ref{eq:ttt-equation})-(\ref{eq:zzz-equation}). For $\eta=1.5$ and
for the depicted initial conditions a numerical integration of these
equations yields the world line reproduced in this figure.}
\label{fig:lattice potential}
\end{figure}
%%%%%%%%%%%%%%%%%%%%%%%%%%%%%%%%%%

The optical master equations are a local expression of
%\begin{align*}
$\delta\int n(t,z)\sqrt{dt^2 -dz^2}=0~,$
%\end{align*}
which is what in Euclidean space corresponds to Fermat's principle of 
least time,
%\begin{align*}
$\delta\int n(x^1,x^2,x^3)\sqrt{(dx^1)^2+(dx^2)^2+(dx^3)^2}=0~.$
%\end{align*}

All world lines are determined by the optical master
Eq.(\ref{eq:ttt-equation})-(\ref{eq:zzz-equation}). A cursory
examination reveals that Eq.(\ref{eq:integral in terms of longitudinal
  time}) together with Eq.(\ref{eq:ttt-equation})
(resp. Eq.(\ref{eq:zzz-equation}) ) implies Eq.(\ref{eq:zzz-equation}) 
(resp. Eq.(\ref{eq:ttt-equation}) ). Consequently, one readily finds
that the world lines satisfy the following single differential equation,
\begin{align}
\frac{d^2z}{dt^2}=-\left[ 1- \left( \frac{dz}{dt}\right)^2\right]
 \frac{1}{2n^2} \left[ \frac{\partial n^2}{\partial t}\frac{dz}{dt} +
 \frac{\partial n^2}{\partial z} \right]
\label{eq:master equation}
\end{align}
Given any laser-induced refractive index, Eq.(\ref{eq:refractive index}),
this \emph{master equation} governs all properties of all possible electron
trajectories resulting from a generic plane wave laser environment.

One of the properties of a spacetime trajectory is how it gets bent
(a.k.a. accelerated) by the refractive index gradient. The right hand
side of Eq.(\ref{eq:master equation}) is proportional to the
directional derivative along the co-moving $z$-axis. Indeed, given that,
in light of Eq.(\ref{eq:integral in terms of longitudinal time}),  the
unit tangent to the 2-d world line is $\left(
\frac{dt}{d\tilde\tau},\frac{dz}{d\tilde\tau}\right)$, the unit vector
along the co-moving $z$-axis is $\left(
\frac{dz}{d\tilde\tau},\frac{dt}{d\tilde\tau}\right)\equiv \hat z$.
It follows that 
\begin{align}
\frac{d}{d\tilde\tau} \left(\frac{dz}{dt}\right)
&=-\left[ 1- \left( \frac{dz}{dt}\right)^2\right] \partial_{\hat z} \ln n
\end{align}
where $\partial_{\hat z} \ln n=
\frac{dz}{d\tilde \tau}\partial_t \ln n  + 
      \frac{dt}{d\tilde \tau}\partial_z \ln n $.
Thus, if $n$ increases along the positive (resp. negative)
$z$-direction, the charge will accelerate into the negative
(resp. positive) $z$-direction.  In other words, opposite to the
Euclidean circumstance, \emph{regions of higher Lorentzian index of
refraction accelerate charges into a direction where the index is
smaller.} A charge gets expelled.

\update{
One manifests an instance of particular interest by stipulating the vanishing of the particle's conserved transverse momentum: $(p_x,p_y)=(0,0)$. Under this stipulation, Eq.(\ref{eq:refractive index}) indicates that the square of the Lorentzian index is proportional to the field intensity. Hence one may conclude that \emph{charges drift away from regions of locally higher intensity regardless of the signs of the charges}, a well-known phenomenon which is usually attributed to the presence of the so-called ponderomotive force. 
}

A second property arises for a particularly interesting case, a charge
driven by the laser field of a standing
wave\cite{Bauer:1995,Kaplan:2005,Pokrovsky:2005}
linearly polarized along, say, the x-axis. For such a wave the
e.m. vector potential is
\begin{align}
A_x(t,z)&=\frac{E_0}{\omega}\sin \omega t \sin \omega z~,~A_y(t,z)=0~.
\end{align}
and the squared index of refraction is
\begin{align}
n^2(t,z)=1+(p_x/m -\eta\sin\omega t\sin\omega z)^2 ~.
\label{eq:squared index for a standing wave}
\end{align}
Such a refractive index $n(t,z)$ forms a two-dimensional optical
lattice of what in Euclidean space are convex and and concave lenses.
Referring to Figure\footnote{In
  this figure we have set $p_x=0$. This suppresses certain interesting
  features associated with \emph{averaged} motion, which have been
  identified in\cite{Kaplan:2005,Pokrovsky:2005}.}~\ref{fig:lattice potential}, one infers from the
depicted trajectory that there exists a linear stack of of convex
``Lorentzian'' lenses all localized around $z=0$ half way between two
adjacent maxima of the refractive index $n(t,z)$. In
Figure~\ref{fig:lattice potential} they are separated (``vertically'')
by an amount $L=\pi$. These lenses guide the center of the oscillating
trajectory.  Moreover, each lender has a \emph{temporal focal length}. For a
laser beam having a standing wave field of frequency
$\frac{\omega}{2\pi}$, this \emph{focal length} is
\begin{align}
\frac{c}{\pi\omega}\frac{1}{\eta^2}\equiv F
\label{eq:focal length}
\end{align}
It is a property pertaining only to ``paraxial'' trajectories,
i.e. those for which the lab velocity of the charge is
non-relativistic: $\left( \frac{dz}{dt}\right)^2 \ll 1$.

The mathematical reasoning that leads to Eq.(\ref{eq:focal length})
is taken directly from geometrical optics. It starts with
Eq.(\ref{eq:master equation}) combined with Eq.(\ref{eq:squared index
  for a standing wave}). The idea is to calculate for a single lens
\[
\Delta\left(\frac{dz}{dt}\right)\equiv \int_0^{2\pi /\omega} \frac{d^2
  z}{dt^2}dt~,
\]
the change in particle velocity over time $2\pi/\omega$ for
trajectories near $z=0$ that start with $\frac{dz}{dt}=0$.
The result for such paraxial spacetime trajectories is 
\[
\Delta\left(\frac{dz}{dt}\right)= -\frac{z}{F}~,
\]
where $F$, the temporal focal length, is given by Eq.(\ref{eq:focal length}).
%%%%%%%%%%%%%%%%%%%%%%%%%%%%%%%%%%%%%%%%%%%%%%%%%%%%%%%%%%%%%%%%%%%%%%%%
\begin{comment}%%%%%%%%%%%%%%%%%%%%%%%%%%%%%%%%%%%%%%%%%%%%%%%%%%%%%%%
it consists of a periodic array of maxima in the refractive
index.  The world line of the particle is the result of the charge
being pushed away from, i.e. scattering from successive maxima
encountered along its path. The index gradient provides the necessary
force field to do this.  The refractive index isograms, as well as the
associated two-dimensional gradient field (Lorentz) orthogonal to it,
are also depicted in Figure~\ref{fig:lattice potential}\footnote{In
  this figure we have set $p_x=0$. This suppresses certain interesting
  features associated with \emph{averaged} motion, which have been
  identified in\cite{Kaplan:2005,Pokrovsky:2005}.}.
\end{comment}%%%%%%%%%%%%%%%%%%%%%%%%%%%%%%%%%%%%%%%%%%%%%%%%%%%
%%%%%%%%%%%%%%%%%%%%%%%%%%%%%%%%%%%%%%%%%%%%%%%%%%%%%%%%%%%%%%%

\emph{Stable and Unstable Motion} -- The existence of a focal length
and the periodic structure encountered by a paraxial particle
trajectory direct attention to the possibility of parametric
instability in its oscillatory motion. Recall that from geometrical optics
one knows that if neighboring convex lenses of a periodic stack are
separated by the same amount $L$, then such a lens system  accommodates
linearly stable paraxial trajectories if and only if \cite{Marcuse:1972,Yariv:1975} 
\[
0<\frac{(Separation~length)}{(Focal~length)}\equiv \frac{L}{F}< 4
\]
From Figure \ref{fig:lattice potential} one sees that the separation
between consecutive lenses is 
\[
L=\pi\frac{c}{\omega}
\]
In light of Eq.(\ref{eq:focal length}) one finds that paraxial
spacetime trajectories oscillate around $z=0$, i.e. are linearly stable if and
only if the laser impulse parameter satisfies
\begin{align}
\eta^2<\frac{4}{\pi^2}~,
\label{stability criterion}
\end{align}
which is to say, the laser intensity should satisfy
$I_{aver}<.56 \cdot (10,000$\AA$/\lambda)^2 \times 10^{18}$watts/cm$^2$. If this inequality is
violated one has a possible type of parametric resonance, which we
alluded near the beginning of this article. \update{For a differential geometric and more detailed analysis of how such a parametric resonance may lead to the breakdown of the ponderomotive approach, one is referred to \cite{Kwa:2012}.}

\begin{figure}[h]
\centering
\epsfig{file=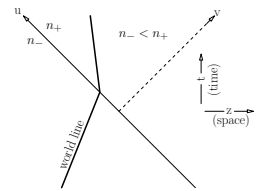,scale=.2}
\caption{Refraction of the world line of a charge}
\label{fig:Refraction of a worldline}
\end{figure}

\emph{Snell's Law}\cite{Kwa:2009} -- A fundamental manifestation of a
refraction process is the manner in which a ray propagates across the
boundary between two regions having different indices of
refraction. In Euclidean space this propagation is expressed by
Snell's law. What is its form for a spacetime medium whose index of
refraction varies (in the limit) discontinuously, as in Figure
\ref{fig:Refraction of a worldline}?

Consider the world line of a charge as it leaves a spacetime medium with
index $n_-$ and enters another one with index $n_+$. Fig. \ref{fig:Refraction of a worldline} gives a close-up view. One introduces
the ``null'' coordinates 
\begin{align}
u&=t-z \quad \textrm{``retarded time'' coordinate}\\
v&=t+z \quad \textrm{``advanced time'' coordinate}
\end{align}
and lets the boundary be the locus of events where
\begin{align}
v=0 \quad \textrm{``history of pulse discontinuity''}~.
\end{align}
It is along characteristics like this where solutions to the wave equation
\begin{align}
\frac{\partial^2}{\partial u\partial v} \left(
                  \begin{array}{c}
                    A_x(u,v)\\
                    A_y(u,v)
                  \end{array}
                                       \right) =0
\end{align}
are allowed to be discontinuous, and hence where the index of
refraction, Eq. (\ref{eq:refractive index}), is allowed
to be discontinuous.

A charge which crosses such a boundary will experience a refractive
index of the form
\begin{align}
n(u,v)=\left\{ \begin{array}{cc}
                   n_-&~v<0\\
                   n_+&~0<v
               \end{array}  
       \right.
\end{align}
The indices are different but constant on either side of the boundary.

The problem is indicated in Figure \ref{fig:Refraction of a worldline}.
One must establish the relationship between the slopes 
\begin{align}
\frac{dv}{du}=\frac{d(t+z)}{d(t-z)}=\frac{1+dz/dt}{1-dz/dt}
\end{align}
on either side of the ``null'' boundary (``characteristic'' of the
wave equation) $v=0$. The solution consists of the statement that
\begin{align}
\left. \frac{dv}{du} n^2\right|_{+}
                  = \left. \frac{dv}{du} n^2\right|_{-} ~,
                          \label{eq:slope law for left travelling pulse}
\end{align}
where the ``+'' and ``-'' refer to $0<v$ and $v<0$ respectively.
This is the Lorentzian version of what is Snell's law in Euclidean space.

The slope $dv/du$ is well known. It is the square of the \emph{Doppler
  factor (``rapidity factor'')} $e^\theta$ for a particle with
$z$-velocity 
$
\frac{dz}{dt}=\tanh \theta\equiv(e^\theta-e^{-\theta})/(e^\theta+e^{-\theta})
$.
The quantity $\theta$ is generally known as the particle's
\emph{rapidity}. The differences and similarities between the
Lorentzian and the Euclidean versions of Snell's law become
particularly perspicuous when one introduces this Doppler factor.
In terms of it Snell's law takes the form\cite{Kwa:2009}
\begin{align}
\left. e^\theta n\right|_{v>0}=\left. e^\theta n\right|_{v<0}
\label{eq:Lorentz-Snell law for left travelling pulse}
\end{align}
across a left-traveling pulse ($v=0$).  This is to be contrasted with
the Euclidean version of Snell's law, which, we recall, is
\begin{align}
n_1\sin \theta_1=n_2\sin\theta_2
\end{align}
The validity of Eq. (\ref{eq:slope law for left travelling pulse}),
and hence of Eq. (\ref{eq:Lorentz-Snell law for left travelling pulse}),
follows directly from Eq.(\ref{eq:integral in terms of longitudinal time})
and the optical master Eqs. (\ref{eq:ttt-equation})-(\ref{eq:zzz-equation})
recast in terms of the ``null'' coordinates $u$ and $v$:
\begin{align}
\frac{du}{d\tilde\tau}\frac{dv}{d\tilde\tau}
                   &=1\label{eq:uv-integral in terms of longitudinal time}
\end{align}
\begin{align}
\frac{d}{d\tilde\tau} n\frac{du}{d\tilde\tau}
                   &=2\frac{\partial n(u,v)}{\partial v}
%\label{eq:u-equation}
%\end{align}
%\begin{align}
~\textrm{and}~\frac{d}{d\tilde\tau} n\frac{dv}{d\tilde\tau}
                   &=2\frac{\partial n(u,v)}{\partial u}~.
\label{eq:v-equation}
\end{align}
By stipulating that $\frac{\partial n}{\partial u}=0$ as in Figure
\ref{fig:Refraction of a worldline}, one can replace $\tilde\tau$ with
$v$ as a worldline parameter. With the help of the second equation of
Eq. (\ref{eq:v-equation}), Eq. (\ref{eq:uv-integral in terms of
  longitudinal time}) leads directly to Eq. (\ref{eq:slope law for
  left travelling pulse}).  

%%%%%% Added 10/24/2012
\update{
\emph{Refractive Index as a Measure of Kinetic Energy Gain} -- Recall that the charge kinetic energy is given by
\begin{equation}\label{eq:kinetic energy}
K(\tau)=m\left(\frac{dt}{d\tau}-1\right).
\end{equation}
Thus, along a given trajectory, 
\begin{equation}\label{eq:kinetic energy gain}
\frac{dK}{d\tau}=\frac{m}{2}\frac{\partial}{\partial t}n^2(t,z)
\end{equation}
by Eqs. (\ref{eq:tt-equation}), (\ref{eq:scalar potential}), and (\ref{eq:refractive
 index}). This equation implies that \emph{a charge gains kinetic energy as it 
enters a region of higher refractive index from one where the index is smaller.}
Conversely, the charge loses kinetic energy as it leaves a region
of higher refractive index and enters one where the index is smaller. In other words, \emph{a change in the Lorentzian refractive index is an indicator of energy exchange between the charge and the e.m. field.} 

Furthermore, since regions of higher index accelerate charges into a direction where the index is smaller, it follows that \emph{charges are inclined to lose their kinetic energy
in response to the e.m. field.} 

The relation between the kinetic energy gain and the refractive index is most 
transparent for charges responding to a traveling wave whose vector potential 
depends only on, say, the advanced time coordinate $u=t-z$. In this case, 
Eq. (\ref{eq:kinetic energy gain}) becomes
\begin{equation}
\frac{dK}{du}=\frac{m}{2p_u}\frac{d}{du}n^2,
\end{equation}
where $p_u$ is the conserved value of $du/d\tau$. Consequently, the
kinetic energy gain changes linearly with the increase in the square of the refractive
index. That is,
\begin{equation}\label{eq:linear energy gain}
\Delta K=\frac{m\Delta n^2}{2p_u}.
\end{equation}
%This, together with the Snell's law (\ref{eq:slope law for left travelling pulse}) or 
%(\ref{eq:Lorentz-Snell law for left travelling pulse}), shows that a change in the
% slope $du/dv$ or equivalently in the particle's rapidity $\theta$ suffices to 
%indicate gain or loss of the particle kinetic energy.

\emph{Energy Gain from Interaction with a Chirped Traveling Pulse} -- As an application, consider the particle dynamics in the chirped plane e.m. field which is polarized along the $x$-axis ($A_y\equiv 0$) and whose field function is given by
\begin{equation}\label{eq:chipred field}
\dfrac{dA_x}{d\tilde{u}}=-{\cal E}\cos\left(\varphi_0+\tilde{u}+b\tilde{u}^2\right)g(\tilde{u})
\end{equation}
as in \cite{Galow:2011,Salamin:2012}. Here ${\cal E}$, $\varphi_0$, $b$, and $\tilde{u}$ are the field amplitude, a constant initial phase, and a dimensionless chirp parameter, and the dimensionless retarded time coordinate $\tilde{u}=\omega_0u$, where $\omega_0$ is the frequency of the field at $(t,z)=(0,0)$. Also,
\begin{equation}\label{eq:pulse-shape}
g(\tilde{u})=\exp\left[-\dfrac{(\tilde{u}-4\sigma)^2}{2\sigma^2}\right]
\end{equation}
is a pulse-shape function, where $\sigma$ is related to the pulse duration $\tau$ (full-width at half-maximum) via $\sigma=\omega_0\tau/(2\sqrt{2\ln 2})$. From (\ref{eq:linear energy gain}), one deduces immediately the evolution of the particle kinetic energy in $\tilde{u}$:
\begin{equation}
\Delta K=\dfrac{q^2{\cal E}^2}{2mp_u}\left[\int\,\cos\left(\varphi_0+\tilde{u}+b\tilde{u}^2\right)g(\tilde{u})\,d\tilde{u}\right]^2,
\end{equation}
which matches the calculation in \cite{Galow:2011,Salamin:2012}. Specifically, one sees that the particle energy gain scales linearly with the field intensity as a consequence of the linear relation between the energy gain and the change in the refractive index squared.}

%%%%% April 15, 2013 %%%%%%%%%%%%%%%%%%%%%%%
\update{
\emph{Conclusion} -- With a Lorentzian index of refraction at its
foundation, relativistic particle dynamics driven by generic
plane wave laser radiation of arbitrary intensity has been mathematized in  terms
\begin{enumerate}
\item
of the Lorentzian (``spacetime'') version of
the optical master equations,
\item
of geodesics on a spacetime manifold with in-general non-zero
curvature (``geometrization of laser radiation''),
\item
of Fermat's principle of least time, 
\item
of Snell's law, 
\item
of the focal lengths of Lorentzian lenses, which
make up the periodic spacetime lattice of a standing wave, and 
\item
of the associated stability criterion for particles moving through
this lattice,
\end{enumerate}
the first four of which are mathematically equivalent.\\
Last, but not least, the Lorentzian index of
  refraction is shown to be an indicator of energy transfer
  between the radiation and the particle. 
}

\bibliography{prlKwaGerlachVanWoerkum2}

\begin{thebibliography}{22}
\expandafter\ifx\csname natexlab\endcsname\relax\def\natexlab#1{#1}\fi
\expandafter\ifx\csname bibnamefont\endcsname\relax
  \def\bibnamefont#1{#1}\fi
\expandafter\ifx\csname bibfnamefont\endcsname\relax
  \def\bibfnamefont#1{#1}\fi
\expandafter\ifx\csname citenamefont\endcsname\relax
  \def\citenamefont#1{#1}\fi
\expandafter\ifx\csname url\endcsname\relax
  \def\url#1{\texttt{#1}}\fi
\expandafter\ifx\csname urlprefix\endcsname\relax\def\urlprefix{URL }\fi
\providecommand{\bibinfo}[2]{#2}
\providecommand{\eprint}[2][]{\url{#2}}

\bibitem[{\citenamefont{Landau and Lifschitz}(1969)}]{LandauMechanics}
\bibinfo{author}{\bibfnamefont{L.}~\bibnamefont{Landau}} \bibnamefont{and}
  \bibinfo{author}{\bibfnamefont{E.}~\bibnamefont{Lifschitz}},
  \emph{\bibinfo{title}{Mechanics}} (\bibinfo{publisher}{Pergamon Press Inc.},
  \bibinfo{address}{Elmsford, NY}, \bibinfo{year}{1969}), pp.
  \bibinfo{pages}{93--95}, \bibinfo{edition}{2nd} ed.

\bibitem[{\citenamefont{Kibble}(1966{\natexlab{a}})}]{Kibble:1966a}
\bibinfo{author}{\bibfnamefont{T.}~\bibnamefont{Kibble}},
  \bibinfo{journal}{Phys. Rev. Lett.} \textbf{\bibinfo{volume}{16}},
  \bibinfo{pages}{1054} (\bibinfo{year}{1966}{\natexlab{a}}),
  \urlprefix\url{http://link.aps.org/doi/10.1103/PhysRevLett.16.1054}.

\bibitem[{\citenamefont{Kibble}(1966{\natexlab{b}})}]{Kibble:1966b}
\bibinfo{author}{\bibfnamefont{T.}~\bibnamefont{Kibble}},
  \bibinfo{journal}{Phys. Rev.} \textbf{\bibinfo{volume}{150}},
  \bibinfo{pages}{1060} (\bibinfo{year}{1966}{\natexlab{b}}),
  \urlprefix\url{http://link.aps.org/doi/10.1103/PhysRev.150.1060}.

\bibitem[{\citenamefont{Bauer et~al.}(1995)\citenamefont{Bauer, Mulser, and
  Steeb}}]{Bauer:1995}
\bibinfo{author}{\bibfnamefont{D.}~\bibnamefont{Bauer}},
  \bibinfo{author}{\bibfnamefont{P.}~\bibnamefont{Mulser}}, \bibnamefont{and}
  \bibinfo{author}{\bibfnamefont{W.~H.} \bibnamefont{Steeb}},
  \bibinfo{journal}{Phys. Rev. Lett.} \textbf{\bibinfo{volume}{75}},
  \bibinfo{pages}{4622} (\bibinfo{year}{1995}),
  \urlprefix\url{http://link.aps.org/doi/10.1103/PhysRevLett.75.4622}.

\bibitem[{\citenamefont{Startsev and McKinstrie}(1997)}]{Startsev:1997}
\bibinfo{author}{\bibfnamefont{E.~A.} \bibnamefont{Startsev}} \bibnamefont{and}
  \bibinfo{author}{\bibfnamefont{C.~J.} \bibnamefont{McKinstrie}},
  \bibinfo{journal}{Phys. Rev. E} \textbf{\bibinfo{volume}{55}},
  \bibinfo{pages}{7527} (\bibinfo{year}{1997}),
  \urlprefix\url{http://link.aps.org/doi/10.1103/PhysRevE.55.7527}.

\bibitem[{\citenamefont{Kaplan and Pokrovsky}(2005)}]{Kaplan:2005}
\bibinfo{author}{\bibfnamefont{A.~E.} \bibnamefont{Kaplan}} \bibnamefont{and}
  \bibinfo{author}{\bibfnamefont{A.~L.} \bibnamefont{Pokrovsky}},
  \bibinfo{journal}{Phys. Rev. Lett.} \textbf{\bibinfo{volume}{95}},
  \bibinfo{pages}{053601} (\bibinfo{year}{2005}),
  \urlprefix\url{http://link.aps.org/doi/10.1103/PhysRevLett.95.053601}.

\bibitem[{\citenamefont{Pokrovsky and Kaplan}(2005)}]{Pokrovsky:2005}
\bibinfo{author}{\bibfnamefont{A.~L.} \bibnamefont{Pokrovsky}}
  \bibnamefont{and} \bibinfo{author}{\bibfnamefont{A.~E.}
  \bibnamefont{Kaplan}}, \bibinfo{journal}{Phys. Rev. A}
  \textbf{\bibinfo{volume}{72}}, \bibinfo{pages}{043401}
  (\bibinfo{year}{2005}),
  \urlprefix\url{http://link.aps.org/doi/10.1103/PhysRevA.72.043401}.

\bibitem[{\citenamefont{Arnold and Avez}(1968)}]{ArnoldAvez:1968}
\bibinfo{author}{\bibfnamefont{V.~I.} \bibnamefont{Arnold}} \bibnamefont{and}
  \bibinfo{author}{\bibfnamefont{A.}~\bibnamefont{Avez}},
  \emph{\bibinfo{title}{Ergodic Problems of Classical Mechanics}}
  (\bibinfo{publisher}{W.A. Benjamin, Inc.}, \bibinfo{address}{New York, NY},
  \bibinfo{year}{1968}), pp. \bibinfo{pages}{81--109}.

\bibitem[{\citenamefont{Misner et~al.}(1973)\citenamefont{Misner, Thorne, and
  Wheeler}}]{MTW:3.1}
\bibinfo{author}{\bibfnamefont{C.~W.} \bibnamefont{Misner}},
  \bibinfo{author}{\bibfnamefont{K.~S.} \bibnamefont{Thorne}},
  \bibnamefont{and} \bibinfo{author}{\bibfnamefont{J.~A.}
  \bibnamefont{Wheeler}}, \emph{\bibinfo{title}{Gravitation}}
  (\bibinfo{publisher}{W.H. Freeman and Company}, \bibinfo{address}{San
  Francisco}, \bibinfo{year}{1973}), chap. \bibinfo{chapter}{3.1}.

\bibitem[{\citenamefont{Jackson}(1975)}]{Jackson:1975}
\bibinfo{author}{\bibfnamefont{J.}~\bibnamefont{Jackson}},
  \emph{\bibinfo{title}{Classical Electrodynamics}} (\bibinfo{publisher}{John
  Wiley and Sons, Inc.}, \bibinfo{address}{New York, NY},
  \bibinfo{year}{1975}), pp. \bibinfo{pages}{551,572}, \bibinfo{edition}{2nd}
  ed.

\bibitem[{\citenamefont{Born and Wolf}(1980{\natexlab{a}})}]{Wolf:1980}
\bibinfo{author}{\bibfnamefont{M.}~\bibnamefont{Born}} \bibnamefont{and}
  \bibinfo{author}{\bibfnamefont{E.}~\bibnamefont{Wolf}},
  \emph{\bibinfo{title}{Principles of Optics}} (\bibinfo{publisher}{Pergamon
  Press Inc.}, \bibinfo{address}{Elmsford, NY},
  \bibinfo{year}{1980}{\natexlab{a}}), p. \bibinfo{pages}{112},
  \bibinfo{edition}{sixth} ed.

\bibitem[{\citenamefont{Kwa}(2012)}]{Kwa:2012}
\bibinfo{author}{\bibfnamefont{K.~H.} \bibnamefont{Kwa}}, \bibinfo{journal}{J.
  Phys. A-Math. Theor.} \textbf{\bibinfo{volume}{45}}, \bibinfo{pages}{105203}
  (\bibinfo{year}{2012}).

\bibitem[{\citenamefont{Kwa}(2009)}]{Kwa:2009}
\bibinfo{author}{\bibfnamefont{K.~H.} \bibnamefont{Kwa}}, Ph.D. thesis,
  \bibinfo{school}{Dept. of Mathematics, O.S.U.}, \bibinfo{address}{Dept. of
  Mathematics, O.S.U., Columbus, OH} (\bibinfo{year}{2009}),
  \urlprefix\url{http://rave.ohiolink.edu/etdc/view?acc_num=osu1250103994}.

\bibitem[{\citenamefont{Brunel}(1987)}]{Brunel:1987}
\bibinfo{author}{\bibfnamefont{F.}~\bibnamefont{Brunel}},
  \bibinfo{journal}{Phys. Rev. Lett.} \textbf{\bibinfo{volume}{59}},
  \bibinfo{pages}{52} (\bibinfo{year}{1987}).

\bibitem[{\citenamefont{Brunel}(1988)}]{Brunel:1988}
\bibinfo{author}{\bibfnamefont{F.}~\bibnamefont{Brunel}},
  \bibinfo{journal}{Phys. Fluids} \textbf{\bibinfo{volume}{31}},
  \bibinfo{pages}{2714} (\bibinfo{year}{1988}).

\bibitem[{\citenamefont{Geindre et~al.}(2010)\citenamefont{Geindre,
  Marjoribanks, and Audebert}}]{Geidre:2010}
\bibinfo{author}{\bibfnamefont{J.~P.} \bibnamefont{Geindre}},
  \bibinfo{author}{\bibfnamefont{R.~S.} \bibnamefont{Marjoribanks}},
  \bibnamefont{and} \bibinfo{author}{\bibfnamefont{P.}~\bibnamefont{Audebert}},
  \bibinfo{journal}{Phys. Rev. Lett.} \textbf{\bibinfo{volume}{104}},
  \bibinfo{pages}{135001} (\bibinfo{year}{2010}).

\bibitem[{\citenamefont{Veltcheva et~al.}(2012)\citenamefont{Veltcheva, Borot,
  Thaury, Malvache, Lefebvre, Flacco, Lopez-Martens, and
  Malka}}]{Velcheva:2012}
\bibinfo{author}{\bibfnamefont{M.}~\bibnamefont{Veltcheva}},
  \bibinfo{author}{\bibfnamefont{A.}~\bibnamefont{Borot}},
  \bibinfo{author}{\bibfnamefont{C.}~\bibnamefont{Thaury}},
  \bibinfo{author}{\bibfnamefont{A.}~\bibnamefont{Malvache}},
  \bibinfo{author}{\bibfnamefont{E.}~\bibnamefont{Lefebvre}},
  \bibinfo{author}{\bibfnamefont{A.}~\bibnamefont{Flacco}},
  \bibinfo{author}{\bibfnamefont{R.}~\bibnamefont{Lopez-Martens}},
  \bibnamefont{and} \bibinfo{author}{\bibfnamefont{V.}~\bibnamefont{Malka}},
  \bibinfo{journal}{Phys. Rev. Lett.} \textbf{\bibinfo{volume}{108}},
  \bibinfo{pages}{075004} (\bibinfo{year}{2012}).

\bibitem[{\citenamefont{Born and Wolf}(1980{\natexlab{b}})}]{BornAndWolf:1980}
\bibinfo{author}{\bibfnamefont{M.}~\bibnamefont{Born}} \bibnamefont{and}
  \bibinfo{author}{\bibfnamefont{E.}~\bibnamefont{Wolf}},
  \emph{\bibinfo{title}{Principles of Optics}} (\bibinfo{publisher}{Pergamon
  Press, Inc.}, \bibinfo{address}{Elmsford, NY},
  \bibinfo{year}{1980}{\natexlab{b}}), chap. \bibinfo{chapter}{3.2},
  \bibinfo{edition}{6th} ed.

\bibitem[{\citenamefont{Marcuse}(1972)}]{Marcuse:1972}
\bibinfo{author}{\bibfnamefont{D.}~\bibnamefont{Marcuse}},
  \emph{\bibinfo{title}{Light Transmission Optics}} (\bibinfo{publisher}{Van
  Norstrand Reinhold Inc.}, \bibinfo{address}{New York}, \bibinfo{year}{1972}),
  pp. \bibinfo{pages}{174--184}.

\bibitem[{\citenamefont{Yariv}(1975)}]{Yariv:1975}
\bibinfo{author}{\bibfnamefont{A.}~\bibnamefont{Yariv}},
  \emph{\bibinfo{title}{Quantum Electronics}} (\bibinfo{publisher}{John Wiley
  and Sons, Inc.}, \bibinfo{address}{New York}, \bibinfo{year}{1975}), pp.
  \bibinfo{pages}{99--104}, \bibinfo{edition}{2nd} ed.

\bibitem[{\citenamefont{Galow et~al.}(2011)\citenamefont{Galow, Salamin,
  Liseykina, Harman, and Keitel}}]{Galow:2011}
\bibinfo{author}{\bibfnamefont{B.~J.} \bibnamefont{Galow}},
  \bibinfo{author}{\bibfnamefont{Y.~I.} \bibnamefont{Salamin}},
  \bibinfo{author}{\bibfnamefont{T.~V.} \bibnamefont{Liseykina}},
  \bibinfo{author}{\bibfnamefont{Z.}~\bibnamefont{Harman}}, \bibnamefont{and}
  \bibinfo{author}{\bibfnamefont{C.~H.} \bibnamefont{Keitel}},
  \bibinfo{journal}{Phys. Rev. Lett.} \textbf{\bibinfo{volume}{107}},
  \bibinfo{pages}{185002} (\bibinfo{year}{2011}),
  \urlprefix\url{http://link.aps.org/doi/10.1103/PhysRevLett.107.185002}.

\bibitem[{\citenamefont{Salamin}(2012)}]{Salamin:2012}
\bibinfo{author}{\bibfnamefont{Y.~I.} \bibnamefont{Salamin}},
  \bibinfo{journal}{Phys. Lett. A} \textbf{\bibinfo{volume}{376}},
  \bibinfo{pages}{2442} (\bibinfo{year}{2012}),
  \urlprefix\url{http://dx.doi.org/10.1016/j.physleta.2012.06.020}.

\end{thebibliography}
\end{document}